\documentclass{ws-procs10x7}

\begin{document}

\title{Branching fractions and CP asymmetries 
in $B\rightarrow h^{+}h^{'-}$}

\author{
Giovanni Punzi\\ for the CDF II Collaboration}

\address{Scuola Normale
Superiore and Istituto Nazionale di Fisica Nucleare\\
Via F. Buonarroti 2, Pisa, I-56100 Italy
\\ E-mail: giovanni.punzi@pi.infn.it
}

\twocolumn[\maketitle \abstract{Relative 
branching fractions of $B_{d,s}\rightarrow h^{+}h^{'-}$ decays (where 
$h,h'= K$ or $\pi$)  and direct CP asymmetry in $B_d\rightarrow K^+\pi^-$ have been measured with 
180 pb$^{-1}$ of data collected by the CDFII detector at the Tevatron 
collider. 
This includes the first BR measurement of a charmless $B_{s}\rightarrow $PP
decay
($B_s\rightarrow K^+K^-$).
}]

\thispagestyle{empty}

\section{Introduction\label{sec:Intro}}

The comparison of measurements performed on a number of such modes at $e^{+}e^{-}$ 
experiments with theoretical expectations has provided a wealth of
 information.

The CDF II experiment\cite{CDF} has access to a large yield of these decays at the Tevatron collider, 
thanks to a dedicated trigger on secondary decay vertices
. In addition, it is also sensitive to the corresponding decays
of $B_s$ mesons and beauty baryons, which provide much additional valuable
information\cite{strangeB,Burasetal}. In particular, 
combining measurements of $B_{s}$ modes with measurements on the $B_u$/$B_d$ sector allows eliminating or
constraining uncertain hadronic parameters.

In this paper we report on the analysis of 2-body charmless decays of $B_d$, $B_s$ and
$\Lambda_{b}$ in a
sample of 179~$\pm$~11~pb$^{-1}$ collected with
the CDF II detector between February 2002 and September 2003.

\section{\label{sec:Data} Sample Selection}

The data sample was selected in the 3--level CDF trigger system 
by a set of requirements dedicated to $B\rightarrow h^{+}h^{'-}$ 
candidates. Two oppositely--charged tracks are required, with $p_T
>2.0$~GeV/c, total $p_T
>5.5$~GeV/c,  impact parameters larger than 100~$\mu$m, and azimuthal opening angle between
$20^{\circ}$ and 135$^{\circ}$.
The $B$ candidate is then required to have a  transverse
decay length larger than 200~$\mu$m, invariant mass between 4.0 and
7.0~GeV/c$^2$, and to point back to the primary vertex within $140\ \mu
m$.

In offline analysis, after reconfirming the trigger cuts with offline
quantities, further cuts are imposed on transverse momenta and impact parameters of the
two tracks, and on transverse decay length, impact parameter and
isolation of the $B$ candidate.
Isolation is defined as $\frac{p_T(B)}{p_T(B)+\sum p_T}$, where the sum 
runs on every other
track within a cone of radius 1 in $\eta$-$\phi$ space around the $B$ 
candidate flight direction. This cut is particularly
effective in rejecting combinatoric background.
The set of cuts was chosen to
maximize the quantity $S/(S+B)^{1/2}$, where $S$ is the number of signal events 
expected from detailed simulation, 
and $B$ is the background estimated by extrapolating the 
sidebands of the data.

\begin{figure}
\epsfxsize160pt
\figurebox{}{}{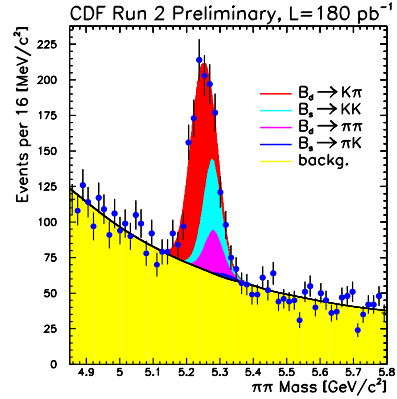}
\caption{$M_{\pi\pi}$ distribution of the $B\rightarrow h^{+}h^{'-}$ 
candidates after all selection cuts. The result of the fit is overlaid.\label{fig:b0hh}}
\end{figure}

The resulting distribution of invariant
mass, with an arbitrarily chosen pion mass assignment to
both tracks, shows a clear signal of $893 \pm 47$ 
events and $\sigma = 38\pm 2$~MeV/c$^{2}$, with a peak S/B in excess of 2 (Fig.~\ref{fig:b0hh}).

Detailed simulation predicts sizeable, closely spaced signals in this 
mass region from two $B_d$
modes: $B_d\rightarrow\pi^+\pi^-$, $B_d\rightarrow K^+\pi^-$ and two
$B_s$ modes: $B_s\rightarrow K^-\pi^+$, $B_s\rightarrow K^+K^-$, 
forming a single unresolved bump.  In addition, the 
$\Lambda_{b}\rightarrow p\pi^{-},pK^{-}$ modes might appear as a slight excess
around $5.5$~GeV/c$^{2}$.

\section{\label{sec:fit} Measurement of individual modes}

The relative contributions to the signal of each $B_{d,s}\rightarrow h^{+}h^{'-}$
component, and the CP asymmetry of the self--tagging $B_d\rightarrow K^+\pi^-$
mode were measured by means of an unbinned likelihood fit
which combines kinematics and PID information on the two tracks.
The Likelihood for the $i^{th}$ event is written as:
\begin{eqnarray*}
\mathcal{L}_i = b\cdot \mathcal{L}^{bck} + (1-b)\cdot \mathcal{L}^{sig} 
\end{eqnarray*}
The index $sig$ ($bck$) labels the contribution of
 signal (background),  and $b$ is the background fraction.
 The signal likelihood function is 
 $\mathcal{L}^{sig} = \sum_{j} f_j\cdot \mathcal{L}^{kin}_j\cdot
 \mathcal{L}^{PID}_j$, where
the index $j$ runs over all possible $B_{d,s}\rightarrow h^+h^{'-}$
modes and the parameters $f_j$ are their relative fractions to be determined by the
fit. 

$\mathcal{L}^{kin}_j$ is in principle a function of the invariant mass of the track
pair, with a  mode--dependent mass assignment to each outgoing particle.
In a rigorous approach\cite{punzibias}, one would then need to write every term in the likelihood 
(including the background part) as functions of the joint (correlated) distributionof all four possible 
masses.

\begin{figure}
\epsfxsize110pt
\figurebox{}{}{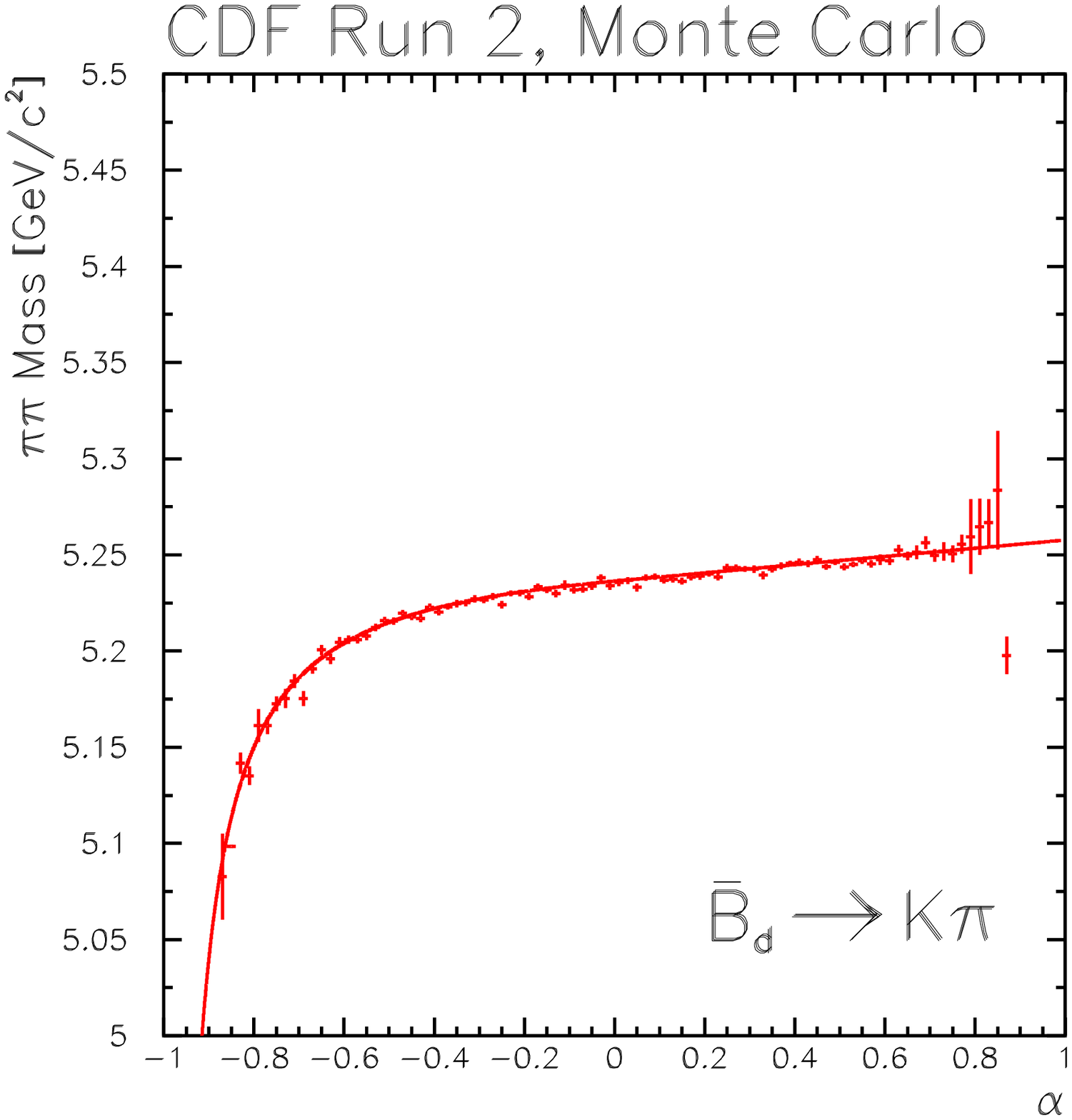}\hspace{-1.1cm}
\epsfxsize110pt
\figurebox{}{}{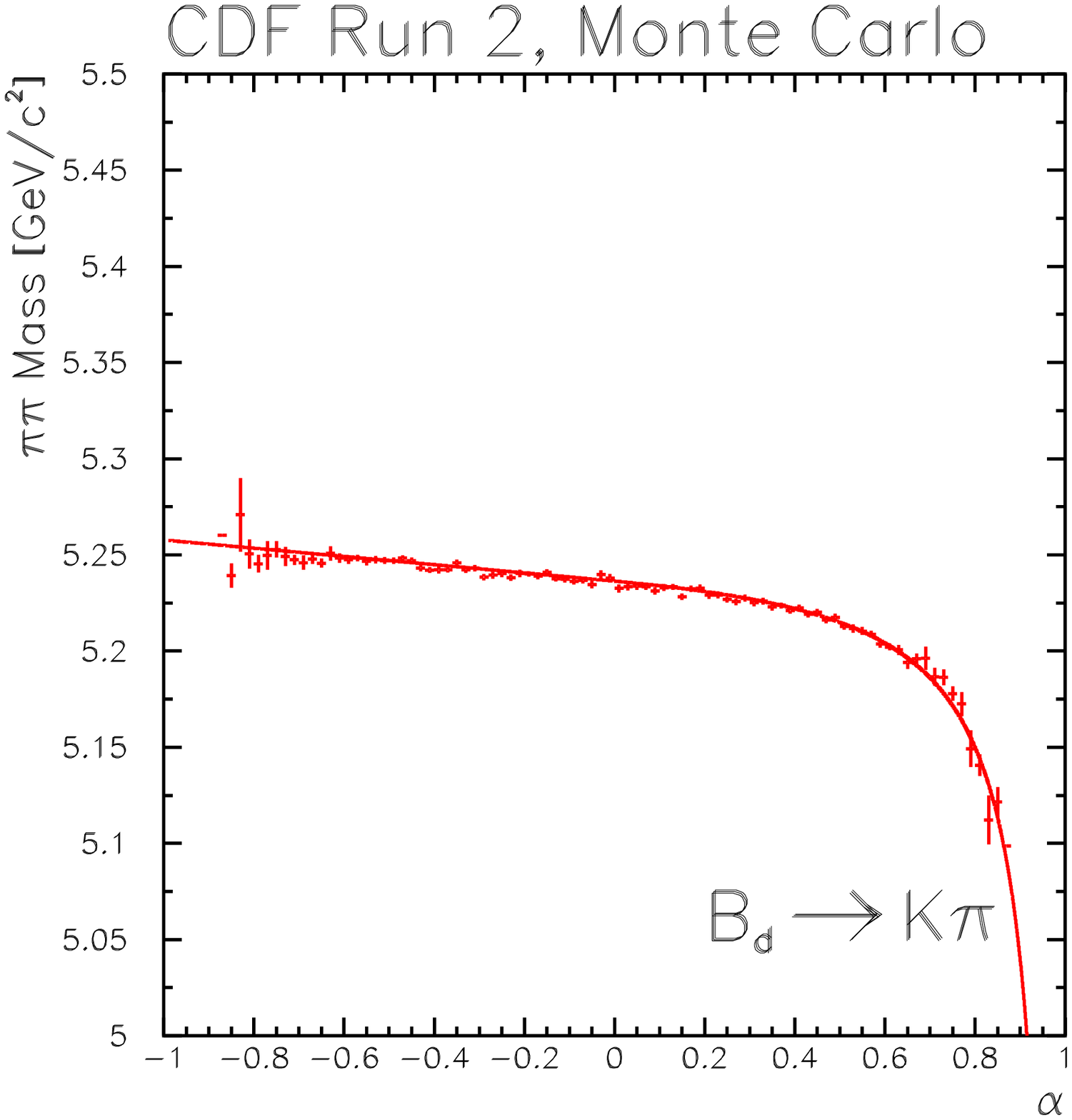}
\caption{(Left) $M_{\pi\pi}$ vs $\alpha$ for $\overline{B}_{d}\rightarrow
\pi^{+}K^{-}$,
compared to the first--order expression: 
$M_{\pi\pi}^{2} = M^2_{B^0_d}+(2-\alpha)(m^2_{\pi}-m^2_K)$. 
(Right) Same for $B_{d}\rightarrow \pi^{-}K^{+}$.
\label{fig:Mpipivsalpha}}
\end{figure}

In order to simplify the problem, only two variables are used: (1) the invariant mass $M_{\pi\pi}$ computed 
with the pion mass assignment to both tracks and (2) a signed momentum imbalance, defined as
$\alpha = (1-\frac{p_1}{p_2}) \cdot q_{1}$, where  $p_1$ ($p_2$) is the  
smaller (larger) of the track momenta, and $q_1$ is the charge of the smaller momentum track.
It can be shown that the candidate mass calculated with any possible mass assignment to the two 
tracks can be written to a good approximation as functions of just $M_{\pi\pi}$ and $\alpha$,
which therefore provide a compact summary of all available kinematic
information (Fig.~\ref{fig:Mpipivsalpha}).

The kinematic term is then written as:
\begin{eqnarray*}\mathcal{L}_j^{kin}=
\frac{1}{\sigma\sqrt{2\pi}}
e^{-\frac{1}{2}\left(\frac{M_{\pi\pi}-\mathcal{M}_{j}(\alpha)}{\sigma}\right)^2} \cdot P_{j}(\alpha)
\end{eqnarray*} 

where the $\mathcal{M}_{j}(\alpha)$ are very simple 
analytical expressions obtained by series expansion; detailed detector simulation
shows that they are very accurate within the kinematic range of interest
(Fig.~\ref{fig:Mpipivsalpha}).
$P_{j}(\alpha)$ is distribution of $\alpha$
for each signal mode after the effect of reconstruction cuts,
as determined from full simulation and parameterized by a $6^{th}$
order polynomial.
The mass resolution $\sigma$ 
is set by rescaling simulation predictions to match 
the observed widths 
of several other two-body decays ($D^0\rightarrow h^{+}h^{'-}$,
$J/\psi \rightarrow \mu^+\mu^-$ and $\Upsilon \rightarrow \mu^+\mu^-$). 
The uncertainty in this rescaling is accounted for in the final systematics.

The Likelihood term related to particle identification information is
\begin{eqnarray*}
\mathcal{L}^{PID}_j&=&
pdf^{PID}_j (\textsf{ID}_{1},\textsf{ID}_{2}, \sigma_{1},\sigma_{2}) 
\end{eqnarray*}
with the observable $\textsf{ID}$ carrying information from the
specific energy release of the track in the drift chamber gas
$dE/dx_{meas}$, and is defined as:
$\mathsf{ID} = 
\frac{\frac{dE}{dx}_{meas}-\frac{dE}{dx}_{exp-\pi}}
{\frac{dE}{dx}_{exp-K}-\frac{dE}{dx}_{exp-\pi}}$ 
 where $dE/dx_{exp-K}$ ($dE/dx_{exp-\pi}$)
is the expected value  in the kaon (pion) hypothesis. The
expected value of $\mathsf{ID}$ is then by construction equal to 0 for a pion, 1 for a kaon.
This parameterization allows using a single observable for both the
pion and kaon terms.

The function $pdf^{PID}$ is a non--factorizable distribution 
of two correlated Gaussians, with track-dependent resolutions
$\sigma_{1},\sigma_{2}$, and a fixed correlation $\sigma_{12}$,
determined from data. The distributions of resolutions have been
checked to be equal for all modes\cite{punzibias}.
The response of the drift chamber has been calibrated with a large sample
of $D^0\rightarrow K \pi$ decays from the same trigger, 
with their sign tagged by the presence of a $D^{*\pm}$. 
The correlation between the measured $dE/dx$ of two tracks in the same
event, due to time--dependent fluctuations of the drift chamber
gain, has been measured with the same sample. 
The average $K/\pi$ separation was measured to be 1.4$\sigma$ in the 
kinematical range of interest. This moderate resolution is however
sufficient to provide a statistical separation between K and $\pi$
which is $60 \%$ of what a perfect PID would provide.
The uncertainty on the calibrations, and the effect of small unmodeled non--Gaussian tails
have also been evaluated from data and included in the final systematic uncertainty.


The background is described by $\mathcal{L}^{bck} = \mathcal{L}^{kin}_{bck} \cdot
\mathcal{L}^{PID}_{bck}$, with the kinematic term: 
\begin{eqnarray*}\mathcal{L}^{kin}_{bck} =
P_{bck}(\alpha)\cdot (e^{c_0+c_1\cdot M_{\pi\pi}}+c_2)
\end{eqnarray*}
where $P_{bck}(\alpha)$ is the distribution of $\alpha$
for background events, obtained from sidebands of real data and
parameterized as a $6^{th}$ degree polynomial.
The  $c_i$ are free parameters in the fit. A few alternative parameterizations 
of the background mass spectrum have been tested, and the corresponding parameter variations included in the
systematics.

The PID term for the background is similar to the signal, and assumes that only pions and kaons are
present, with each track having an independent probability $f_{\pi}$ to be a pion.
The pion fractions $f_{\pi}$ are left free in the fit to vary independently in three mass regions 
(left, under, and right of the signal peak).

The complete likelihood fit has been tested on MonteCarlo samples, and 
showed gaussian pulls on all 
variables with unit sigma and negligible biases.

\section{\label{sec:fitresults}Efficiency corrections}

To extract relative branching fractions from the raw fit results it is necessary to 
correct for different efficiency of the selection on each decay. 
Most corrections are obtained from detailed detector simulation; exceptions are 
the efficiency of the $B$ isolation cut (only affecting $B_{d}/B_{s}$ 
ratios), the difference in efficiency 
of the trigger track processor for kaons and pions due to their different energy
release in the drift chamber gas, and the charge asymmetry. 
These three effects are not reliably simulated, and have been measured 
from real data.
A systematic has been added  for possible
differences between $B_d$ and $B_s$ momentum spectra\cite{mangano},
that are treated as equal in the simulation. 

Isolation efficieny has been measured as a function of $B$ transverse 
momentum from fully reconstructed samples of $B_s\rightarrow J/\psi
\phi$, $B_s\rightarrow D_{s}\pi$,  $B_d\rightarrow J/\psi K^{*}$, $B_d\rightarrow 
D^{-}\pi^{+}$. 
The difference in trigger efficiency for pions and kaons has been measured
from an unbiased sample of pions and kaons from charm decays.
The momentum dependent charge asymmetry of CDF tracking has been measured
on large  samples of inclusive tracks and on $K_{s}\rightarrow \pi^{+}\pi^{-}$
decays\cite{D0paper}, and used to correct our measurement of direct $A_{\mathsf{CP}}$.
The size of this correction is about $1\%$ and its relative uncertainty
($\pm25$~\%) is included in the final systematics.

The resulting total efficiency varies between modes by less then
$12\%$.

The total systematic uncertainties are dominated by 
uncertainties in $dE/dx$ calibration and efficiency of the isolation
cut. Both uncertainties are due to statistical uncertainties in the
calibrations samples, and are therefore expected to decrease with
sample size.

\section{\label{sec:results}Results}
The ratio of the $B_{d}$ modes:
\begin{eqnarray*}
 \frac{BR(B_d\rightarrow\pi^\pm\pi^\mp)}{BR(B_d\rightarrow K^\pm\pi^\mp)} =
 0.24\pm 0.06 \pm 0.05
\end{eqnarray*}
is in good agreement with the current world average\cite{spkr:HFAG}: $0.25\pm 0.025$.
The direct CP asymmetry is:
\begin{eqnarray*}
\frac{N(\overline{B}^0_{d}\rightarrow K^-\pi^+) 
- N(B^0_{d}\rightarrow K^+\pi^-)}{N(\overline{B}^0_{d}\rightarrow K^-\pi^+) 
+ N(B^0_{d}\rightarrow K^+\pi^-)} 
=\\
= -0.04 \pm 0.08 \pm 0.006
\end{eqnarray*}
which is compatible both with zero, and with the measurements performed by
BaBar and Belle collaborations\cite{ACP-BaBelle}.

For the $B_{s}$ sector:
\begin{eqnarray*}
\frac{f_s \cdot BR(B_s\rightarrow K^\pm K^\mp)}
{f_d\cdot BR(B_d\rightarrow K^\pm\pi^\mp)} =
0.50 \pm 0.08\pm 0.07
\end{eqnarray*}
Since this is a CP eigenstate, it is possible for its lifetime to be
different from the average $B_{s}$ lifetime measured from semileptonic
modes. The above result is obtained under the assumption that
$\Gamma_s=\Gamma_d$,
the $B_s\rightarrow K^+K^-$ mode is dominated by the short--lived
component , and that $\Delta\Gamma_s/\Gamma_s = 0.12\pm0.06$ (the latter
uncertainty is included in the quoted systematics). 
We also quote:
\begin{eqnarray*}
\frac{f_d \cdot BR(B_d\rightarrow \pi^\pm\pi^\mp)}
{f_s\cdot BR(B_s\rightarrow K^\pm K^\mp)} =
0.48 \pm 0.12\pm 0.07
\end{eqnarray*}

From the above numbers, the absolute BR can be
obtained\cite{spkr:PDG2004}:
\begin{eqnarray*}
 BR(B_s\rightarrow K^\pm K^\mp) =
 34.3 \pm 5.5 \pm 5.2 \cdot 10^{-6}
\end{eqnarray*}
This is almost twice the BR of $B_d\rightarrow K^\pm\pi^\mp$ which differs only in the
spectator quark, in agreement with recent calculations of flavor--SU(3) breaking\cite{Khodjamirian,Burasetal}.
This result will hopefully be a useful input for theoretical
models of charmless decays.
The large yield ($\approx 230$ events in current sample) also offers an interesting
opportunity for a measurement of $\Delta\Gamma_{s}$ without the need
for an angular analysis.

No evidence is found for the other $B_{s}$ decay for which a sizeable
BR is expected, and a limit\cite{F-C} is set:
\begin{eqnarray*}
\frac{f_s\cdot BR(B_s\rightarrow K^\pm \pi^\mp)}
{f_d\cdot BR(B_d\rightarrow K^\pm \pi^\mp)} <
0.11~@~90\%~C.L.
\end{eqnarray*}
This translates to\cite{spkr:PDG2004}:
\begin{eqnarray*}
 BR(B_s\rightarrow K^\pm \pi^\mp) < 7.5 \cdot 10^{-6}~@~90\%~C.L.
\end{eqnarray*}
which is close to the lower end of current expectations.\cite{}
In addition to the above, rarer modes (BR: $10^{-8}$ to $10^{-7}$) dominated by penguin annihilation and 
exchange diagrams have been searched for by adding their expected contributions to the Likelihood. 
In each case, the fit parameters changed by a
negligible amount, and no evidence was found for those modes. 
The following limits\cite{F-C} are set:
\begin{eqnarray*}
\frac{BR(B_d\rightarrow K^\pm K^\mp)}
{BR(B_d\rightarrow K^\pm \pi^\mp)} <
0.17~@~90\%~C.L.
\end{eqnarray*}
or\cite{spkr:PDG2004}:
\begin{eqnarray*}
BR(B_d\rightarrow K^\pm K^\mp)< 3.1\cdot 10^{-6}~@~90\%~C.L.
\end{eqnarray*}
The current best limit in this mode is\cite{HFAG} 
$0.6\cdot 10^{-6}$.
\begin{eqnarray*}
\frac{BR(B_s\rightarrow \pi^\pm \pi^\mp)}
{BR(B_s\rightarrow K^\pm K^\mp)} <
0.10~@~90\%~C.L.
\end{eqnarray*}
This is derived under the assumption that 
both modes have the same average lifetime.
This translates\cite{spkr:PDG2004} to:
\begin{eqnarray*}
BR(B_s\rightarrow \pi^\pm \pi^\mp)<3.4\cdot 10^{-6}~@~90\%~C.L.
\end{eqnarray*}
which is a substantial improvement over the previous best limit\cite{HFAG}
 of $1.7\cdot 10^{-3}$.

The same data have also been used in a separate search for the charmless decay 
$\Lambda_b^0\to p\pi^-, pK^-$, by counting the number of events found in the 
search window  $5.415 - 5.535$ GeV/c$^{2}$. The choice of the window and 
the cuts have been optimized for maximum sensitivity\cite{punziopt}.
No evidence for signal is found, and a Bayesian upper limit, based 
on uniform prior, is obtained at 90\% credibility: 
$BR(\Lambda_b^0\to p\pi^- /pK^-)\leq 22\times10^{-6}$.

This is an improvement over the previous limit:
$BR(\Lambda_b^0\to p\pi^-, pK^-)\leq 50\times10^{-6}$ at 90\%CL~\cite{Buskulic:1996tx}.
Predictions\cite{Mohanta:2000nk} lie in the range $2-3\times10^{-6}$.
Improved sensitivity is expected in the future with inclusion of PID
information for the proton.

\section*{Acknowledgments}
The author wishes to thank R.~Fleischer and J.~Matias for helpful
discussions.

\end{document}